# CREW HaT: A Magnetic Shielding System for Space Habitats

Desiati, Paolo[1]  
*Wisconsin IceCube Particle Astrophysics Center (WIPAC) & Physics Department*  
*University of Wisconsin – Madison, Madison, WI 53706, U.S.A.*

*and*

D'Onghia, Elena[2]  
*Department of Astronomy & Physics Department*  
*University of Wisconsin – Madison, Madison, WI 53706, U.S.A.*

**At the dawn of a new space exploration age, aiming to send humans back to the Moon and for the first time to Mars, it is necessary to devise a solution to mitigate the impact that space radiation has on spacecraft and astronauts. Although technically challenging, active magnetic shielding is generally considered a promising solution. We propose a lightweight deployable system producing an open magnetic field around a space habitat. Our Cosmic Radiation Extended Warding (CREW) system consists of a cylindrical Halbach array coil arrangement, or Halbach Torus (HaT). This configuration generates an enhanced external magnetic field while suppressing it in the habitat volume. The CREW HaT takes advantage of recent innovations in high-temperature superconductors (e.g., ReBCO) that enables the needed high currents. We present a preliminary feasibility design of the magnetic shielding system and its collapsible mechanical structure to sustain the internal magnetic forces while protecting astronauts. We also lay down the next steps towards a more evolved and comprehensive device design.**

## Nomenclature

*CL* = confidence level  
*GCR* = galactic cosmic rays  
*ISS* = international space station  
*ReBCO* = rare-earth barium copper oxide  
*SPE* = solar particle events  
*Sv* = Sievert  

## I. Background

THE 21st century is the age of human space exploration. NASA's priority is to send humans back to the Moon[1] soon and eventually travel to Mars in the next decade. However, human expeditions will be possible only when we find a solution to mitigate the cosmic radiation hazards on spacecraft and astronauts (as shown on Ref. 2). Space radiation in cislunar and interplanetary space consists of solar particle events (SPE) and galactic cosmic rays (GCR). SPE particles, primarily electrons and light ions with energy up to about 100 MeV/nucleon, are originated from solar flares and coronal mass ejections. The burst events can last from hours to days with very intense particle flux. On the other hand, GCR particles consist of a steady and relatively low flux of particles, primarily protons and heavy ions with energy of 1 GeV/nucleon and above, and constitute a significant risk factor for long-duration missions such as those to Mars. Space radiation particles can penetrate habitats, spacecraft, equipment, and spacesuits and harm astronauts. Minimizing the physiological changes caused by space radiation exposure is one of the biggest challenges in keeping astronauts fit and healthy as they travel through the solar system. Ionizing radiation is a serious problem that can cause damage to all parts of the body, including the central nervous system, skin, gastrointestinal tract, skeletal system, and blood-forming organs.

---

[1] Research Scientist, WIPAC, 222 W. Washington Ave #500, Madison, WI 53703, U.S.A. (desiati@wipac.wisc.edu)  
[2] Associate Professor, Dept. of Astronomy, 475 N. Charter St, Madison, WI 53706, U.S.A. (edonghia@astro.wisc.edu)

Astronauts on the International Space Station (ISS) receive radiation doses from space between 80 and 160 mSv per 6-month mission.[3] In comparison, a standard medical chest X-ray delivers approximately 0.1 mSv. Although inconclusively, there is a 3% excess risk of death from cancer at a cumulative dose of 1000 mSv,[4,5] just below the 1200 mSv dose expected from a trip to Mars. Recently, NASA recommended that habitat systems have sufficient protection to reduce by 15% the exposure to the GCRs compared to free space. This achievement would keep the absorbed radiation dose from GCRs below 1.3 mSv per day for habitats in space.[6] This risk factor sets the standard for setting radiation limits for men and women astronauts at different ages using estimates above the 95% confidence level (CL) for uncertainties in risk projection models.[5] Uncertainties occur related to predicting particle energy spectra, and the limited understanding of heavy-ion radiobiology leads to a level of uncertainty that requires extra margin when setting radiation limits.

The eleven-year solar cycle modulates the exposure to GCR particles, with minimum exposure occurring during solar maximum when high-speed solar wind pushes away lower energy GCRs from the solar system. Although exposure to GCRs can be reduced by approximately a factor of two by timing Mars mission during solar maximum, solar cycles are hard to predict. The corresponding increase in SPE occurrences and intensity needs mitigation technologies and poses the question of exploring new active activity shielding technology.[6]

There are two different approaches to protect spacecraft and astronauts from the harmful effects of space radiation: passive and active shielding. With passive shielding, space radiation is absorbed through a layer of material. It is the current solution adopted in space. Any material can stop ionizing particles up to a certain energy, above which they penetrate the shield.[7] For instance, an aluminum slab with 2, 4, 10 cm thickness (corresponding to a mass thickness of approximately 5, 11, 27 $g/cm^2$, respectively, which is the product of the density, 2.7 $g/cm^3$, and thickness) can stop protons up to about 50, 70, 125 MeV.[8] Passive shielding may work for high-Z GCR particles at energies below 1 GeV/nucleon. However, while losing energy penetrating the habitat's walls, GCR particles induce nuclear reactions, generating lower energy secondary particles, including gamma rays and neutrons.[9] Therefore, astronauts inside the habitat are exposed to the risk of this internal mixed radiation environment. Another problem is the excessive weight needed to achieve acceptable radiation mitigation. Polyethylene is a good shielding material because it has high hydrogen content, and hydrogen atoms efficiently absorb radiation.[7,10-17] However, it may not be the best solution for a long-time space mission (e.g., Mars).[18,19] An example of passive shielding is the one suggested for the Artemis mission to ensure the safety of future space crews traveling to the Moon by producing well-fitted vests to wear in space.[20,21] While low-cost and lightweight, the vests are designed to protect vital human organs from radiation. Although this solution might work for the softer SPE particles, it has negligible protection from the more energetic GCR and secondary debris. As a result, it is inadequate to shield humans in long-haul travels to Mars.

The most promising approach to protect astronauts from cosmic radiation consists of generating a magnetic field that surrounds the spacecraft, similar to the Earth's magnetosphere. In fact, on Earth, the geomagnetic field acts as an extended cocoon that shields life on the surface against the harmful space radiation. Feasibility studies showed that this strategy is superior to any passive shielding approach in terms of efficiency in reducing the impact of radiation on spacecraft. However, mass and power consumption are critical to the practical realization of such future devices.[7,22-26] Compared to the large and heavy confined magnetic field systems studied in the past,[7,22,23] structurally transparent and open concepts, more similar to the Earth's magnetic field, provide a lighter-weight solution with a significantly lower impact surface for secondary radiation production.[7] The challenge for an open magnetic field geometry is to achieve the best compromise among mechanical structure weight, strength to support the intense magnetic forces, and astronauts' safety. A deployable system combined with the open configuration for the magnetic field would be ideal, providing the additional appealing aspect of facilitating transport in reduced volumes before initiating full operations.

## II. The CREW HaT concept

The Cosmic Radiation Extended Warding (CREW) magnetic shield concept is based on a cylindrical Halbach array arrangement, or Halbach Torus (HaT). In the embodiment described here, the system consists of eight electromagnet racetrack coils disposed around the habitat region with rotating magnetic polarities (shown on the left of Figure 1). Such a configuration produces a magnetic field extending in the space around the CREW HaT while suppressing it in the habitat region at the center. Each coil is equipped with several rare-earth barium copper oxide (ReBCO) high-temperature superconducting tape wrappings to achieve the needed currents and generate a magnetic field in the surrounding space. The coils are equipped with dedicated cryocoolers, thus avoiding single-point failures. Solar panels may be considered to power the cryogenic pumps independently. Each coils' containment structure is designed to support hoop stresses from self-induced magnetic forces. In addition, the coil support structure to the main body is designed to be deployable and withstand inter-coil magnetic forces.

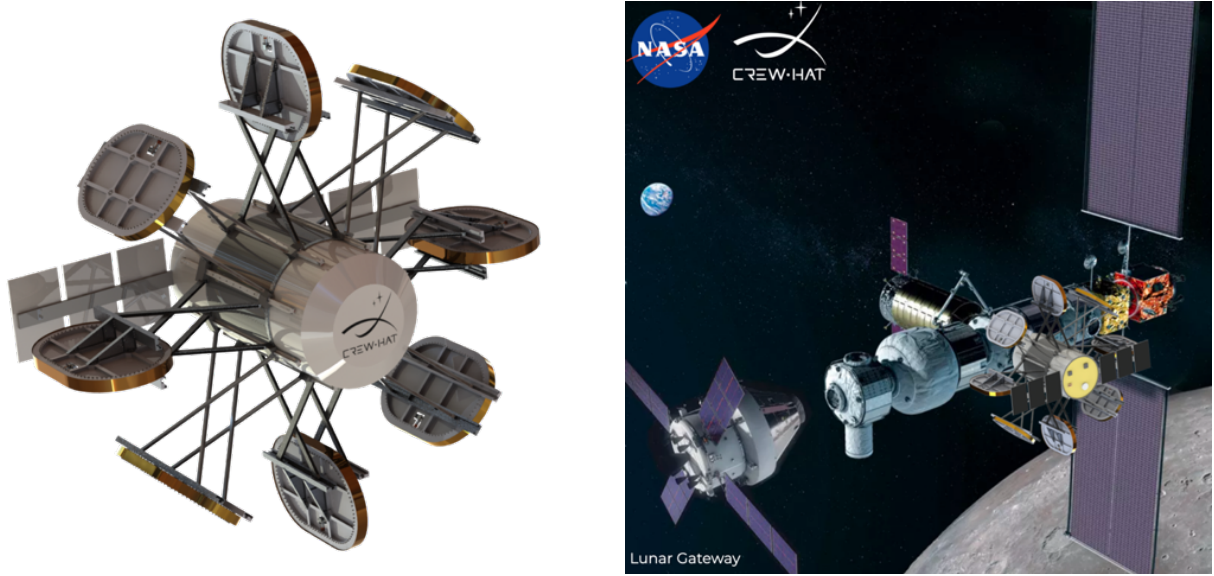

**Figure 1. Design rendering of the CREW HaT magnetic shield device on the left. Artist's rendering of the Lunar Gateway's HALO module with the docked CREW HaT magnetic shield on the right (Lunar Gateway rendering courtesy of NASA).**

The design process started with an optimization procedure of the system's geometrical configuration aiming at minimizing the number of radiation particles reaching the habitat region while reducing the magnetic forces impinging on the coils. Such a constraint aims to reduce the overall system's weight. The preliminary design applied safety factors between 1.5 and 2 to the various mechanical components to achieve an adequate margin that allows a buffer within operational conditions for material failures.

The CREW HaT magnetic shield is meant to be constructed on Earth and transported to space in a folded configuration. Once docked to the habitat, the deployment mechanism activates the scissor lift arms to place and lock the coils in their final operational position. On the right of Figure 1, an artist's rendering shows the CREW HaT system docked on a Lunar Gateway habitat.

## III. Feasibility Study

The feasibility study of CREW HaT begins with an iterative process where the system's geometrical configuration and current are varied within an interval of values, limited by the overall constraints imposed by the Halbach torus configuration and by the maximum mechanical stresses we want to achieve.

### A. Geometry Optimization

The core component of the geometry optimization procedure is the numerical integration of particle trajectories in magnetic fields (see Ref. 27 for more details about the code used for the calculations). The equation of motion of charged particles in magnetic fields (expressing the Lorentz Force)

3
International Conference on Environmental Systems

$$\vec{F} = Ze\,\vec{v} \times \vec{B}, \tag{1}$$

with *Ze* the GCR particle's electric charge, $\vec{v}$ its velocity and $\vec{B}$ the magnetic field, is numerically solved using the fourth-order Runge-Kutta integration method. Adaptive time-stepping is used to ensure that sufficiently high numerical accuracy is preserved during the calculations. The magnetic field vector generated by the CREW HaT's coils (shown on the left of Figure 2) is calculated at each point in space where the particle is located.

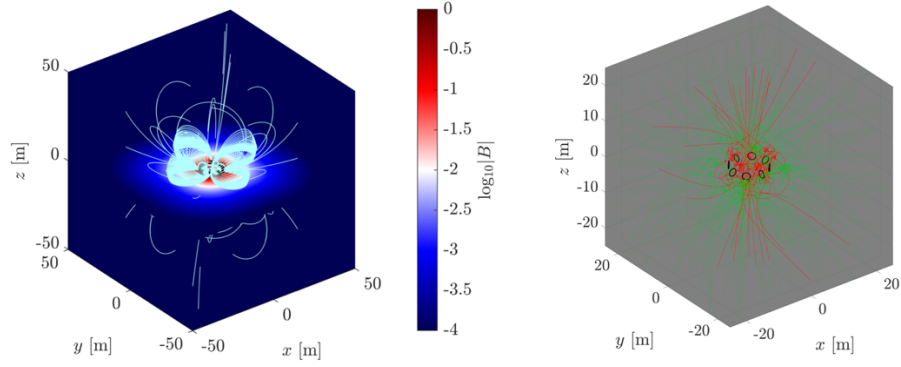

**Figure 2. Visualization of the magnetic field generated by the CREW HaT (on the left). The magnetic field B is expressed in units of Tesla. Visualization of the integrated proton trajectories passing through the CREW HaT's magnetic field (on the right). The trajectories effectively diverted away from the astronauts' habitat are in green. The trajectories that hit the habitat are in red.**

Several proton trajectories are integrated using the CREW HaT's magnetic field. The protons' initial positions are randomly distributed on a sphere with a radius larger than the magnetic field's influence distance and directions towards the central habitat region. Their energy scans the range 1-1000 MeV at seven logarithmically distant intervals (i.e., 1, 3, 10, 30, 100, 300, 1000 MeV). With the magnetic field generated by such a coils' geometry, particles propagating towards the end caps of the Halbach Torus have an easy way through (red trajectories on the right of Figure 2), unlike those propagating perpendicularly to it (green trajectories in the figure). This is similar to the GCR particles penetrating the geomagnetic field on the polar regions, compared to those more effectively diverted back to space from the equatorial zone. The optimization procedure ultimately aims to reduce the number of "red" trajectories as much as possible. The minimum distance of every trajectory from the center of the habitat is determined for each set of values of the geometrical parameters, such as geometry, current configuration (illustrated in Figure 3), and proton energy.

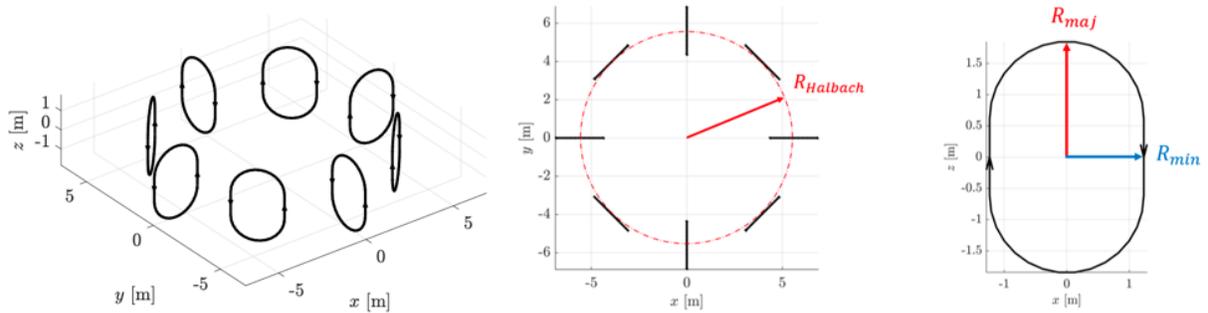

**Figure 3. Geometric parameterization of the Halbach Torus, in isometric view (on the left), top view (on the center), and the illustration of one racetrack coil (on the right).**

| Parameter | Value |
| --- | --- |
| $R_{Halbach}$ | 5.60 m |
| $R_{maj}$ | 1.85 m |
| $R_{min}$ | 1.23 m |
| $I_c$ | $4 \times 10^6$ A |

**Table 1. Optimized geometric parameters of the CREW HaT.**

Any trajectory crossing a sphere at the center with a 3 m radius is assumed to have hit the habitat. The ratio between deflected ($N_d$) and incident ($N_i$) trajectories

$$\eta_s = \frac{N_d}{N_i}, \quad (2)$$

is referred to as shielding efficiency. If $\eta_s = 1$, all particles are deflected, and the shielding efficiency is 100%. The optimization process requires maximizing the shielding efficiency $\eta_s$ within the mentioned geometric constraints. An additional constraint that sets an upper bound to the coils' current $I$ is to minimize the maximum inter-coil magnetic force $F_{max}$, which depends on the square of $I$. At the current stage of our investigation, we decided to target the magnetic shield to particles up to a few hundred MeV to inspect the overall feasibility of such a system. For this reason, the preliminary optimization procedure was performed using 100 MeV energy proton trajectories, and the optimized geometric parameters' values are shown in Table 1. The CREW HaT can deflect about 40% of 100 MeV GCR protons. Because particles react to magnetic fields depending on their energy-to-charge ratio (or E/Z, called rigidity), this is equivalent to the same reduction of GCR He particles at 200 MeV and Fe particles at 2600 MeV. The next step of the feasibility study will explore CREW HaT shielding capabilities at GeV-scale and the possibility of integrating passive shielding at the end-cap regions. It will utilize absorbed radiation dose as the primary optimization metric instead of shielding efficiency.

**B. Magnetic Forces**

The magnetic field generated by an electric current $I$ flowing through an arbitrarily shaped conductor is given by the Biot-Savart Law

$$\vec{B_1} = \oint \frac{\mu_0 I_1}{4\pi} \frac{\vec{dl_1} \times \hat{r}}{r^2}, \quad (3)$$

where the integral is along the length of the conductor, and $r$ is the distance of the current element $\vec{dl_1}$ to the point where we want to evaluate the magnetic field. Any magnetic field interacts with an electric current element $\vec{dl_2}$ flowing through any conductor nearby with the Lorentz force

$$d\vec{F} = I_2 \, \vec{dl_2} \times \vec{B_1}. \quad (4)$$

In multi-coil systems such as the CREW HaT, each coil is subject to magnetic forces induced by its own and neighbor's coils' current. The self-induced forces impinge radially on the coils with a null net force (see left of Figure 4). However, they produce substantial hoop stresses on the coil's wiring and casing, determining the mechanical structure design needed to hold it. In a racetrack coil, the forces are highest at the curved sections where, in the current design, they reach about 900 tons, corresponding to hoop stresses up to about 1 MPa (i.e., 10 atm). Because the magnetic field created by the array at the location of one coil is significantly smaller than the magnetic field each coil produces on itself, intercoil-induced forces are about an order of magnitude lower than the self-induced forces.

The four *tangential* coils facing the CREW HaT's axis of symmetry experience a complex load distribution resembling a *Pringles chip* (a saddle shape also known as hyperbolic paraboloid), resulting in a net outward force of about 85 tons. On the other hand, the four radially positioned coils are subject to radially inward net forces of about 140 tons (see right of Figure 4). Such intercoil forces contribute to additional non-uniform hoop stress on the coils. The overall net force of the CREW HaT is null (assuming all coils are operating under the same conditions). However, the coils' mechanical support structure must sustain compressive and tensile forces during operations.

A preliminary partial failure scenarios study performed to evaluate the extent of the net force balance change should one of the CREW HaT coils fail, shows that the maximum tensile and compressive force on the coils change by less than 10%. However, the study of the resulting shift in net force balance is still in progress.

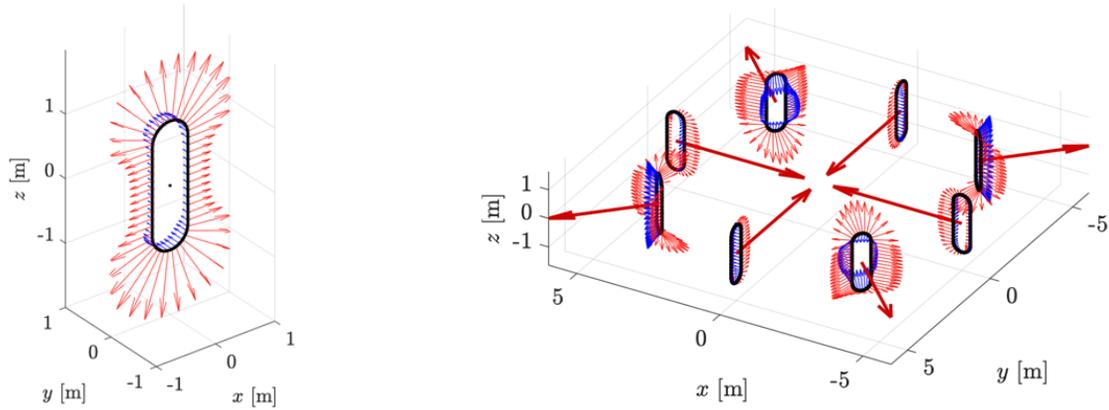

**Figure 4.** *Left*. **Illustration of self-induced magnetic forces on a racetrack coil (in red) with the magnetic field overlayed (in blue).** *Right*. **Illustration of intercoil-induced magnetic forces.**

The calculations show that the maximum intercoil-induced net forces decrease with more elongated racetrack coils. However, the geometry described here is a compromise between lowering such forces while keeping shielding efficiency high.

## IV. Structural Design

The large magnetic forces coupled with the complex loading shown in Figure 4 strongly influence the coil containment structure and the overall CREW HaT mechanical structure design. Additional requirements come from the need for a deployable mechanism that keeps the system folded for launch and expands to its final configuration while maintaining the static operating conditions associated with the superconducting coils.

### C. Design Components

The magnetic forces described in the previous section highlight four different components in the design: the coil containment, the radial support beams, the central support ring, and the folding mechanism. The coil containment has the task of counteracting the radial hoop stresses and the *Pringles chip* folding forces. The beams holding the coils solidly anchored to the central body of the CREW HaT must support against buckling. A dedicated central self-support system must withstand all the compressive and tensile forces and the large moment on the radial beams arising from the alternate compressive and tensile loads. Folding functionality is desirable to reduce the system's volume during transport and reduce bending loads on long support beams during launch.

### D. Coil Containment

The primary role of the coils' containment is to counteract the Lorentz forces induced on the superconductor's windings and prevent them from ripping apart. In addition, it must include thermal insulation to the superconductor and mechanical connections to the CREW HaT structure.

Research conducted for the ITER Fusion Reactor[28] indicates that composite materials such as an epoxy resin-Kevlar mix should provide the necessary strength for the coil containment structure. The preliminary design of the CREW HaT coil structure is based on the studies performed for the Space Radiation Superconducting Shield Project (SR2S).[29] They concluded that Kevlar offers the same strength as titanium with a third of the weight.

The centerpiece of the coil containment is an aluminum alloy ribbed support structure, with two vertical and two horizontal I-beam ribs (as shown in Figure 5), designed to facilitate the superconductor wrapping process. Two C-shaped shells are then slid over and bolted through the ribbed support, providing containment and partial restrain against hoop stresses. The coil containment is then wrapped with a 5 mm layered fiber-reinforced Kevlar composite, which provides the primary restraint against hoop stresses (with at least a factor of two redundancy to reduce possible fiber rupture by stress-induced degradation and, eventually, a sudden structural failure). The stresses induced by the magnetic forces during operations are within the yield strength of the aluminum component, and the strain is below 0.4% required by the wire to maintain its superconducting properties (see section IV.G). With this preliminary design, the mass for one superconducting coil is estimated to be about 2,600 kg. Multi-layered insulation (also called reflectors) maintains the superconducting coil at its cryogenic operating temperature while being irradiated by the Sun's heat.

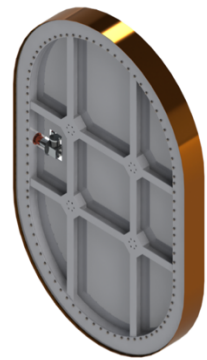

**Figure 5. Coil Containment structure design**

### E. Coil Support Structure

The collapsible coils' support system is based on a *scissor lift* design due to its simplicity and integrity in supporting compressive loads, which have a larger magnitude (see section III.B). Each lifting mechanism consists of four supporting aluminum alloy I-beams with two on each side of a coil. The two pairs of I-beams are pinned together at their midpoint. Each coil's casing is pinned to two sets of trolleys that allow sliding along the I-beams as the structure unfolds. The coils' ribs are used as a mounting structure for U-channel beams on which the scissor lift beams are attached. The tangential coils are directly bolted to the U-channel, whereas the radially oriented coils are mounted using a triangular gusset plate. The bases of the I-beams are then pinned to two additional sets of trolleys that ride along a T-shaped rail attached to a ring structure that surrounds the astronauts' habitat (see Figure 6).

When the structure is folded in the preliminary design, the support I-beams are folded down onto the central ring support. The CREW HaT's unfolding is made possible by the motorized trolleys that expand the scissor lift to its final configuration, where the trolleys lock into position. The I-beams are designed to support the extensive compressive and tensile loading from magnetic forces exerted during operation. The maximum stresses on the beams were estimated to be about half the aluminum yield stress, thus providing a safety margin of two.

Each trolley consists of a C-shaped carriage clasping around the rails. Both the dimension of the trolleys and the rails are designed to withstand contact stresses from the magnetic forces. The unfolding mechanism currently under consideration uses motorized trolleys that slowly spin a pinion gear that meshes with a gear rack etched into the rail surface. The ring structure on which the rails are attached is assumed to be made with aluminum alloy with a 2 cm thickness. The ring is designed to support the compressing load concentrated on the locked locations of each trolley during operation, which will guarantee safety for astronauts in the habitat.

Although the overall mass, as shown in Table 2, is proportional to the loads that the structure must withstand, the current preliminary design is not optimized for minimal mass just yet. The coils account for about 85% of the overall mass, most of which (about 70%) is from the superconducting tape windings. The conductor is a critical component of the CREW HaT shielding optimization. It has to carry sufficient current to produce a magnetic field capable of diverting a significant fraction of space radiation particles. At the same time, it is subject to an upper current level to minimize the loads and stresses from magnetic forces. The actual length and weight of the superconductor may be subject to variations once the optimization refines.

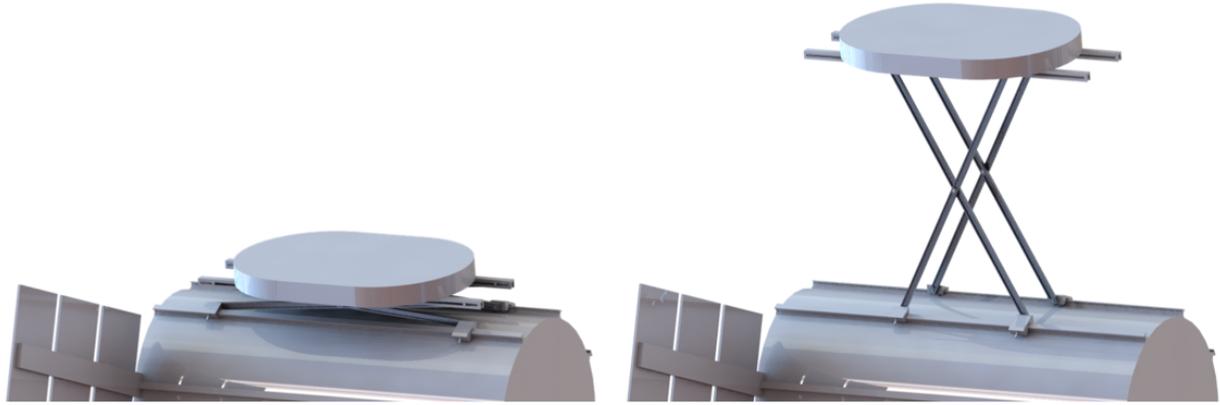

**Figure 6. Rendering of the scissor lift folding mechanism when stowed (on the left) and expanded (on the right).**

| Component | Mass [kg] |
|---|---|
| Eight coils | 2,600 × 8 |
| Ring Structure | 1,371 |
| Scissor Beams | 1.191 |
| Trolleys | 663 |
| **Total Mass [kg]** | **24,025** |

**Table 2. Breakdown of CREW HaT components' mass.**

The central ring structure does not account for a significant fraction of the CREW HaT weight. Nevertheless, it does not need to be made of a solid slab of metal. Since the bulk of the stress is concentrated near the final trolley position on the rails, the midsection may be composed of a lighter material, such as an aluminum honeycomb or Al-composite material. When the CREW HaT is in its unfolded operating configuration, the rails on the central ring will bear the load from magnetic forces.

**F. Deployment Mechanism**

When the structure is folded, in one preliminary design, the coils' orientation is identical to that at operation (see left and center of Figure 7). With this design, the stowed system has a diameter of about 8 m, assuming a 3 m diameter for the ring support. A variation of this design that is currently under consideration involves a rotation mechanism that allows the *radial* coil to collapse and latch onto the *tangential* coil while folded in launch mode. With this option, the CREW HaT stowed diameter is about 6.6 m (right of Figure 7). While the payload fairing for the current Space Launch System (S.L.S.)[30] and SpaceX's Falcon Heavy[31] cannot accommodate this size, the future S.L.S. Block 2 series will have fairings of 8.4 m and 10 m$^3$, and SpaceX Starship will be 9 m with an expected payload of 100-120 tonnes for the Mars mission.[32] On the one hand, CREW HaT's design is still in its preliminary stages, with the overall dimension and size of its components not finalized yet. On the other hand, active magnetic shielding technology is projected for future missions where large and powerful launch systems will be available. Whether the CREW HaT will be launched along with the spacecraft or separately and docked on the spaceship in orbit before the deep-space mission will be studied in detail.

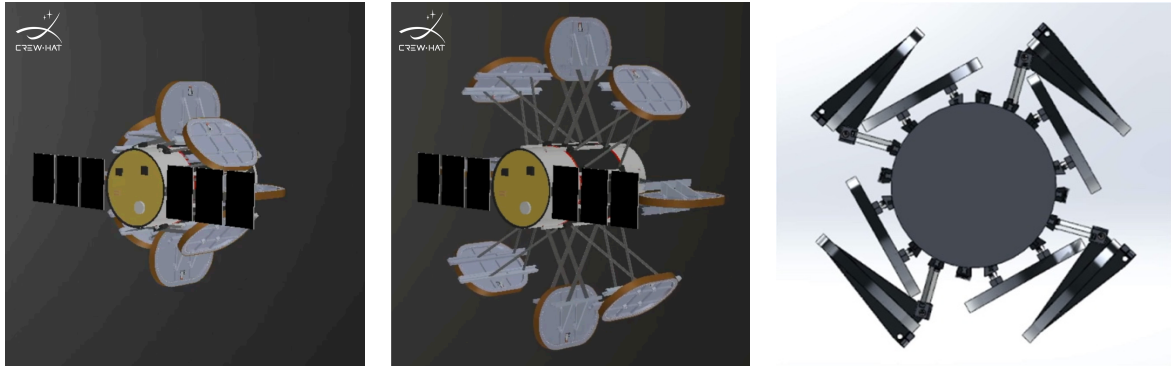

**Figure 7. CREW HaT stowed to its initial folded state (on the left) and fully deployed to its final operating configuration (on the center). Design variation with perpendicular coils' rotation to reduce the stowed size (on the right).**

### G. Superconducting Wire and Cryogenic System Requirements

The preliminary feasibility study, described in section III, shows that the optimized current is $4\times 10^6$ A, corresponding to a maximum magnetic field strength of about 10 T.[33] This current can be achieved by wrapping the conductor $N$ times so that a current $I_c$ is effectively amplified to $N\times I_c$. To achieve such high currents, it is possible to use the recently developed high-temperature superconductors such as ReBCO. These wires' high current carrying capabilities significantly reduce the magnets' weight, while the strength of the fields can be about three times higher than of more conventional superconductors.[34]

The preliminary design uses 4 mm wide, 0.1 m thick 2G HST ReBCO tapes by SuperPower Inc.[35]. A high-purity aluminum tape is wrapped between each strand of ReBCO tape to provide a thermal conduction path to the cryocooler. At an operating temperature of 40 K, a 10 T magnetic field can be achieved and sustained at a critical current $I_c = 100\,A$. To achieve the effective operating current, the superconductor tape is wrapped in a bundle of 210 by 126 mm² cross-sectional area (as shown in Figure 8). The coil containment structure's design (described in section IV.D) must limit the superconducting wire's strain to 0.4% to keep the critical current to its nominal value.[36]

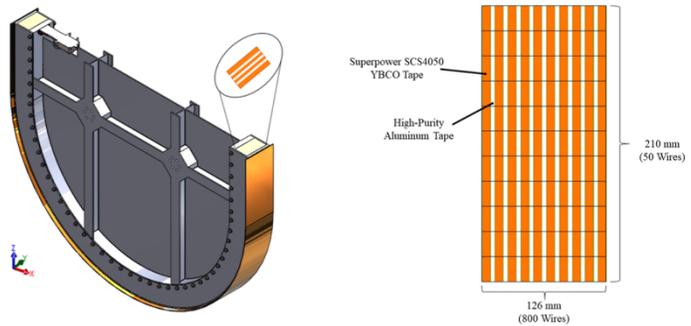

**Figure 8. The superconducting wire bundle cross-section.**

The superconducting wire operating temperature of 40 K poses specific requirements to operate the CREW HaT. Although the advantage of high-temperature superconductors is liquid nitrogen over the more expensive liquid helium, helium gas systems offer lighter cryogenic solutions.[23] There are two primary sources of heat to counteract: solar irradiation, which can be reduced with multilayered reflectors down to about 50 W, and resistive heating, which is estimated to yield less than 1 W. Because of the physical distance between the CREW HaT coils, the preliminary design employs a solution with a small cryocooler on each of the eight coils. The market provides cryocoolers rated at 70 W cooling power operating at 40 K.[37] These devices are light (16 kg) and have an input power of 7.2 kW, for a total of 57.6 kW. Detailed studies will follow how active cooling requirements can be reduced by exploiting the low temperature of space and efficient and uniform insulation from solar irradiation.

To turn CREW HaT to its operational configuration, we must account for the coils' inductance. In the preliminary design and estimation, we may power the coils with eight power supplies (10 V with 100 A maximum current) which would power up the system in about 10 minutes. An alternative could be to use 1 V power supplies with 100 A maximum current, which would power up the system in about 2 hours. These estimates depend on the final optimized coil geometry and operating currents. We expect this to change should we achieve different final configurations as the result of our feasibility studies.

## V. Full-body dose equivalent estimation

Using the CREW HaT magnetic shield design presented here, the full-body dose equivalent absorbed by astronauts inside a spacecraft without and with magnetic shielding was estimated using a similar procedure described in Ref. 29. The GCR flux at solar minimum from Ref. 38 was used for the calculations after comparing it with actual experimental measurements collected on the Cosmic-Ray DataBase (CRDB).[39] Particle propagation and interactions were calculated using the Monte Carlo code GEANT4.[40-42] The full-body dose equivalent calculations were conducted using a computational phantom[43] inside an idealized 1 in thick aluminum spacecraft sphere of 1.6 m radius. For this preliminary calculation, ten million uniformly distributed GCR protons were injected on the phantom only (the astronaut in free space, as a benchmark) and on the phantom inside the aluminum sphere (astronaut inside a proxy spacecraft) without and with the CREW HaT magnetic shield. Figure 8 shows the dose equivalent normalized to free space at three energies.

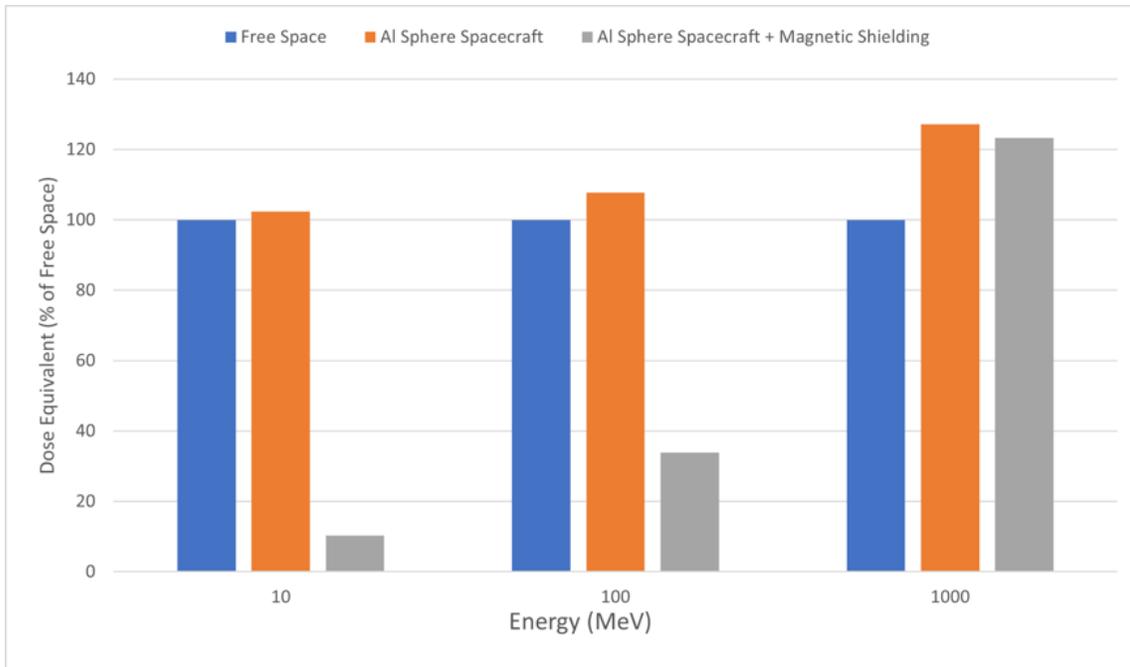

**Figure 9. Preliminary calculation of full-body equivalent dose relative to free space vs. GCR proton energy. Blue bars correspond to astronauts in free space. Orange bars for an astronaut inside a 1 in thick aluminum sphere of radius 1.6 m. Gray bars correspond to the case with CREW HaT operational.**

The presence of a spacecraft and the absence of magnetic shielding increase the dose equivalent due to the secondary particles produced by the impact of the primary GCRs with the spacecraft. When the magnetic shield is turned on, low-energy GCR particles are diverted away, so they don't produce secondary particles and don't contribute to the absorbed dose. As mentioned in section III.A, the preliminary design of the CREW HaT was optimized on the 100 MeV energy scale. The next stage of feasibility studies will optimize the CREW HaT shielding capabilities on the GeV scale, where GCR flux peaks, making it the main requirement for active shielding.

## VI. Conclusions

The idea to devise solutions for radiation protection during human long-haul space exploration was proposed approximately 50 years ago, and dedicated technical studies have been conducted over the last two decades. These pioneering studies provided the notion that active magnetic shielding technology embodies a better option than passive shielding in safeguarding astronauts from space radiation. However, it became clear that the dimension of such magnetic shielding devices, their mass, and the strict requirements for operating in deep space would make their design and construction very challenging. One practical conclusion was that open magnetic field geometries with lower coil density and lighter mechanical structures with smaller architectural mass are options to investigate.[21,22] Not only do these solutions naturally reduce the production of secondary particles, one of the main limitations of the pioneering closed-geometry concepts, but they may provide a better chance to become a reality in the upcoming new space age.

CREW HaT is an open field geometry that provides a lightweight structure that can be folded for transport. The preliminary feasibility study presented here is the first attempt to explore the parameter space of such a magnetic shielding system. The CREW HaT magnetic shield uses the most recent development in high-temperature superconducting technology and is conceived for future long-haul travel in deep space, such as the Mars mission. It is a technology that aims to protect astronauts from the long-term effects of the continuous rain of GCR particles on spacecraft and the debris that penetrate inside the habitats. Future studies will investigate the requirements for reducing radiation dose equivalent well below the career limits and refine the magnetic field geometry that more appropriately translates such requirements into the design and construction of the magnetic shield device using robust and lightweight material and novel folding mechanisms. One of the critical aspects of such studies is to explore the possibility of exploiting the low temperature of space to relax the requirements for active cooling of the superconductive coils, thus saving power. The solution is contingent on technology able to efficiently shield the coils from solar irradiation and guarantee a high degree of uniform temperature to prevent quenching. Another key aspect is to understand how radiation shielding scales with the dimension of the CREW HaT. The benchmark for our future studies is the 9 m diameter Starship spacecraft. Adapting a magnetic shield system to various applications, targeting different spacecraft sizes and shapes, and different particle energies is a crucial aspect to explore.

## Acknowledgments

We thank the senior students of the course of Engineering Mechanics and Aerospace Engineering at the Department of Mechanical Engineering of the University of Wisconsin – Madison for their excellent work: C. Haese, K. Heck, M. Landry, S. Moravec, M. Tuman (EMA 569 spring 2021), X. Halverson, C. Paulman, T. Schewe, M. Walden, M. Baker, R. Fronsee, N. Schneider, M. Wasserman (EMA 469 fall 2021) with whom we closely collaborated on the CREW HaT design. We thank Prof. Sonny Nimityongskul for accepting our project for his senior courses and Prof. John Pfotenhauer for the helpful discussions on the properties of high-temperature superconducting wires and cryogenics. Many thanks to students Sam Garcia and Tyler Fredryck for the equivalent dose calculations and our collaborator Prof. Bryan Bednarz for guiding us in understanding radiation dose and providing access to the computational tools.